# Strain-enabled dynamical magnetoelectric coupling in $BaFe_{12}O_{19}$

ChanJu You[1], Jayanti Higgins[2], Zhiren He[3], Yilin Evan Li[4], Harikrishnan KP[2], Soeren Buchenau[2], Christo Guguschev[5], Mario Brützam[5], Jiaqiang Yan[6], David Mandrus[6], David A. Muller[2,7], Darrell G. Schlom[4,5,7], Guru Khalsa[3], Ankit S. Disa[2*]

[1]Department of Physics, Cornell University, Ithaca, New York 14850, USA
[2]School of Applied and Engineering Physics, Cornell University, Ithaca, New York 14853, USA
[3]Department of Physics, University of North Texas, Denton, Texas 76203, USA
[4]Department of Materials Science and Engineering, Cornell University, Ithaca, NY, USA
[5]Leibniz-Institut für Kristallzüchtung, 12489 Berlin, Germany
[6]Department of Materials Sciences and Engineering, The University of Tennessee, Knoxville, Tennessee, USA
[7]Joint Institute for Neutron Sciences, Oak Ridge National Laboratory, Oak Ridge, Tennessee 37831, USA
[8]Kavli Institute at Cornell for Nanoscale Science, Ithaca, NY, USA
[*]Corresponding author: asd47@cornell.edu

## Abstract

Optical excitation has emerged as a powerful tool for manipulating and creating non-equilibrium phases of matter. Yet, the pathways available for optical control are intrinsically limited by the symmetries and interactions of the equilibrium system. Here, we demonstrate that these pathways can be redesigned through atomic-structural engineering to access hidden dynamical properties. By combining resonant vibrational excitation with epitaxial strain, we realize a light-induced, room-temperature multiferroic state in strained thin films of the technologically relevant hexagonal ferrimagnet, barium hexaferrite ($BaFe_{12}O_{19}$, BaM). Time-resolved second harmonic generation and magneto-optical Kerr effect measurements reveal an enhanced ferroelectric polarization together with a weakening of the magnetic moment upon exciting mid-infrared phonons in strained BaM thin films, but not in the bulk material. Structural analysis and first-principles-based theory show that this effect is enabled by strain-generated displacement of $Fe^{3+}$ ions, which break the local symmetry and facilitate an emergent dynamical magnetoelectric coupling. Both the strong lattice excitation and strain-induced symmetry breaking are essential for enabling this dynamical coupling, promising an important template for designing otherwise inaccessible non-equilibrium phenomena.

## Main

Excitation of materials with ultrashort optical pulses has emerged as a powerful tool to control phases and tune material properties. While many such optical excitations are dominated by non-equilibrium electronic populations and incoherent scattering, more deterministic control has been available through nonlinear optical and phononic effects. These nonlinear pathways allow light-matter interactions dictated by the underlying symmetries and couplings of the crystal, tailoring the system towards desired physical properties. For example, mid-infrared pulses in the THz frequency range can be used to resonantly drive phonons, which couple anharmonically to other modes, including phonons, magnetic and orbital excitations, or strain along particular directions[1-10]. This approach, called 'nonlinear phononics', has led to light-induced ferroelectricity[11,12], melted charge order[13], and manipulation of magnetic states[14-18].

While the results highlight the versatility of this approach for functional control of complex phases, the light-driven states are still limited to those accessible under the symmetry restrictions of the material. For instance, it is not possible to induce arbitrary symmetry-breaking structural distortions if such interactions are not present in the crystal potential. Here, we demonstrate that light-matter interactions can be engineered at the atomic scale to access new dynamically driven effects, which we exploit to realize light-driven magnetoelectricity and multiferroicity at room temperature.

*M*-type barium hexaferrite (BaM) is a well-known high-temperature ferrimagnet ($T_c$ = 723 K) with a large magnetic moment (40 $\mu_B$/unit cell). Its robust magnetic properties, high chemical stability, and low cost of synthesis have made it commercially ubiquitous in various applications involving permanent magnets, information storage, and microwave devices[19]. In addition to its strong magnetism, BaM shows a dramatic softening of its lowest frequency polar phonon and a rapid growth in its dielectric constant which saturates at low temperatures[20-22]. These behaviors indicate proximity to a ferroelectric phase transition, reminiscent of the well-known incipient ferroelectric $SrTiO_3$. BaM, which crystallizes in the magnetoplumbite structure (Fig. 1a) with its magnetization aligned along the *c*-axis[23,24], has $Fe^{3+}$ ions occupying five inequivalent crystallographic sites, among which the trigonal bipyramidal (TBP) site is particularly significant for the material's incipient ferroelectric behavior[20-22, 25]. Experimental studies have suggested that quantum fluctuations of the displacements of $Fe^{3+}$ ions from the TBP mirror plane are responsible for the suppression of long-range ferroelectric order[26-28].

Theoretical calculations have predicted that, as in $SrTiO_3$, the application of sufficiently large strain can stabilize ferroelectricity in BaM at or near room temperature, which would realize a new, robust, multiferroic material[29]. Unfortunately, the required (compressive, in-plane) strain in BaM exceeds 5%, which is inaccessible by conventional approaches, such as pressure, chemical substitution, or thin film epitaxy[30-33].

In addition to strain, one may consider a second control parameter: optical excitation. It was previously shown that a metastable ferroelectric state can be induced optically in $SrTiO_3$ by exploiting the symmetry-allowed nonlinear coupling between a resonantly driven infrared-active phonon ($Q_{IR}$) and strain ($\varepsilon$)[11]. In BaM, we show that, while neither strain nor optical excitation in isolation can stabilize long-ranged multiferroicity due to the symmetry constraints of the crystal, their combination provides a pathway for multiferroic control with light (summarized in Fig. 1a). The introduction of a small epitaxial strain locally breaks the mirror symmetry at the TBP site, creating polar nanoregions but preserving long-range centrosymmetry (Fig. 1b). This local symmetry breaking, however, permits a new magnetoelectric coupling that enables phonon-driven dynamical control of ferroelectric polarization and magnetization with light.

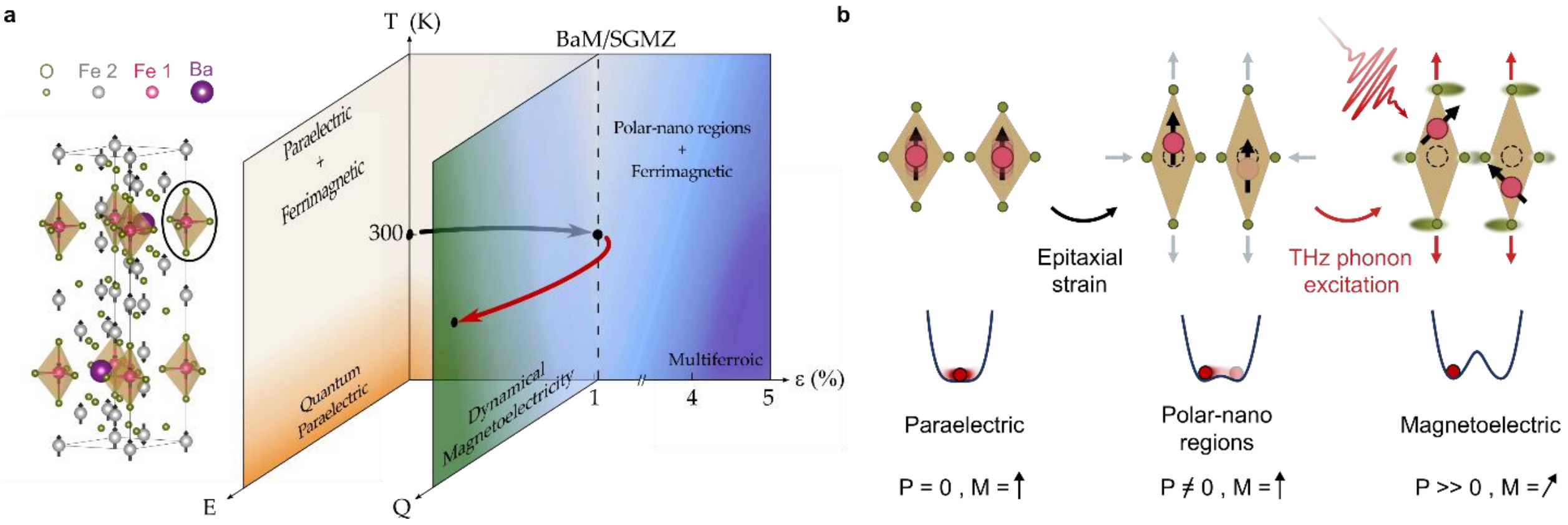

***Figure 1. Strain-enabled dynamical phase map of $BaFe_{12}O_{19}$ (BaM). a)*** *The crystal structure of BaM is shown alongside the phase diagram that shows possible pathways to non-equilibrium phases via different degrees of freedom. The trigonal bipyramidal (TBP) site is marked by the black circle. At room temperature, the system starts at the paraelectric and ferrimagnetic state. With strain, the central $Fe^{3+}$ion displacements break the mirror symmetry at the TBP site, creating polar-nano regions. Phonon excitation from the weakly strained state allows a new coupling mechanism that drives the system into the dynamical magnetoelectric phase.* ***b)*** *Schematic of the proposed pathway to a room-temperature dynamical magnetoelectric state, via local symmetry breaking that enables the coupling of magnetization to polarization in the driven state. In this dynamically driven state, the net polarization is increased while the magnetization is moderately reduced due to slight in-plane canting.*

## Results

### Epitaxial strain and polar nanoregions

In this work, we examine bulk BaM single crystals, as well as unstrained and strained thin films. The films were grown by oxide molecular-beam epitaxy with (0001) surface orientation on sapphire (unstrained) and (Mg,Zr):$SrGa_{12}O_{19}$ (SGMZ)[34], an isostructural substrate with an in-plane lattice constant 1.1% smaller than that of BaM, which imposes a biaxial compressive strain (see Methods). It was previously shown, using multislice electron ptychography, that the epitaxial strain stabilizes polar nanoregions in the film[35,36]. The ptychographic reconstruction shown in Fig. 2a, clearly resolves the off-center displacements of the $Fe^{3+}$ ions at the TBP sites across successive layers along the *a*-axis (white dashed regions). The two spots on either side of the mirror plane indicate the coexistence of positively and negatively off-centered atomic sites with no long-range ferroelectric order[37]. The incident probe polarization dependence of the SHG signal, analyzed with emitted SHG polarization parallel ($P_{out}$) and perpendicular ($S_{\text{out}}$) to the scattering plane, is consistent with the $C_{6v}$ point group expected for the polar phase, reflecting a small net polarity arising from the disordered off-center displacements at the TBP sites, which is absent in the bulk single crystal (Fig. 2b). Meanwhile, the MOKE hysteresis loop shown in Fig. 2c confirms that the ferrimagnetic moment remains robust under applied strain, with a maximum Kerr rotation of approximately 0.7 mrad, in good agreement with previous reports[38-40]. Together, these measurements show that epitaxial strain stabilizes a locally symmetry-broken, magnetically ordered ground state featuring polar nanoregions, which serves as the starting point for the light-driven magnetoelectric coupling described in the subsequent sections.

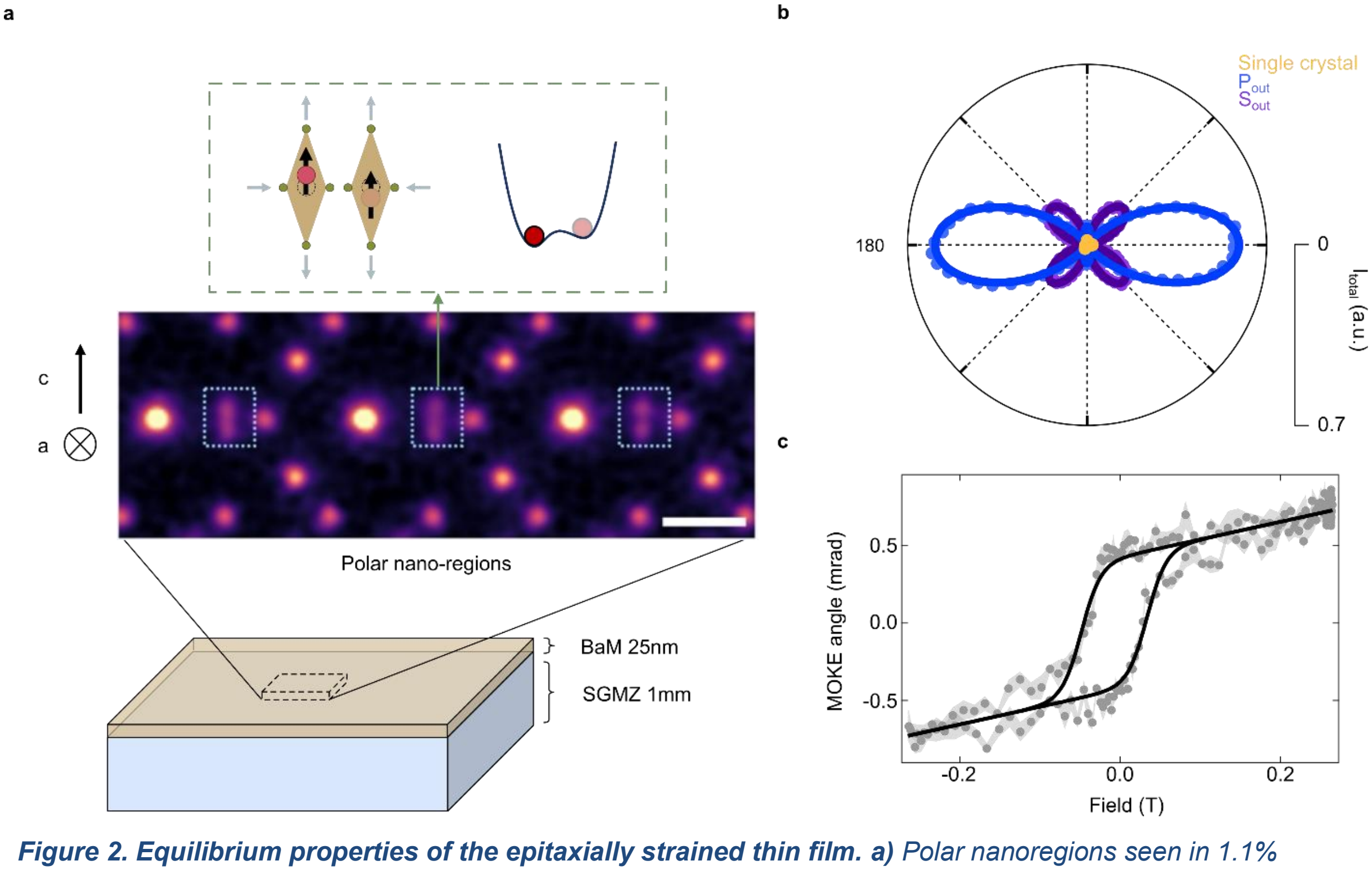

***Figure 2. Equilibrium properties of the epitaxially strained thin film. a)*** *Polar nanoregions seen in 1.1% epitaxially strained BaM film via multislice electron ptychography (Reproduced with permission from Ref. 37). The scale bar in the inset of the ptychography data is 2 Å.* ***b)*** *Static SHG signal of the epitaxially strained film showing the expected $C_{6v}$ point group associated with the polar phase.* ***c)*** *Static MOKE signal of the same strained film.*

## Phonon-selective control of polarization and magnetization

Guided by these experimental observations that compressive strain pushes the system closer to a ferroelectric (and multiferroic) ground state, we aim to stabilize long-range multiferroicity by leveraging the nonlinear phononic coupling to strain, which takes the symmetry-allowed form $V_{NL} = gQ_{IR}^2\varepsilon$, where $g$ is the phonon-dependent coupling constant[11,41]. Based on density functional theory calculations of the coupling strengths (see Supplementary Information), we target the doubly degenerate $E_{1u}$ phonon mode at 17.4 THz. We pump the unstrained and strained BaM films with intense, resonant mid-infrared pulses centered at this frequency with ~400 fs pulse duration and electric field strengths up to 11 MV/cm (see Methods). Time-resolved SHG and MOKE measurements using 800 nm and 400 nm probe pulses, respectively, are used to track the ultrafast evolution of polarization and magnetization following the resonant excitation at room temperature.

The time-resolved SHG and MOKE signals are shown in Figure 3a. In the unstrained case, no change in the SHG signal is observed upon pumping (similar results are found for bulk single crystals, see Supplementary Information). In contrast, the strained film shows a sharp enhancement immediately after photoexcitation, followed by a slower long-lived component. A similar behavior is observed in the pump-probe MOKE response (Fig. 3a, bottom), where only the strained sample exhibits a rapid decrease in Kerr rotation (equivalently, magnetization, see Supplementary Information) of up to 40% that saturates within ~5 ps. These results point toward the broken local symmetry of the strained film enabling new nonlinear optical pathways absent in the unstrained material.

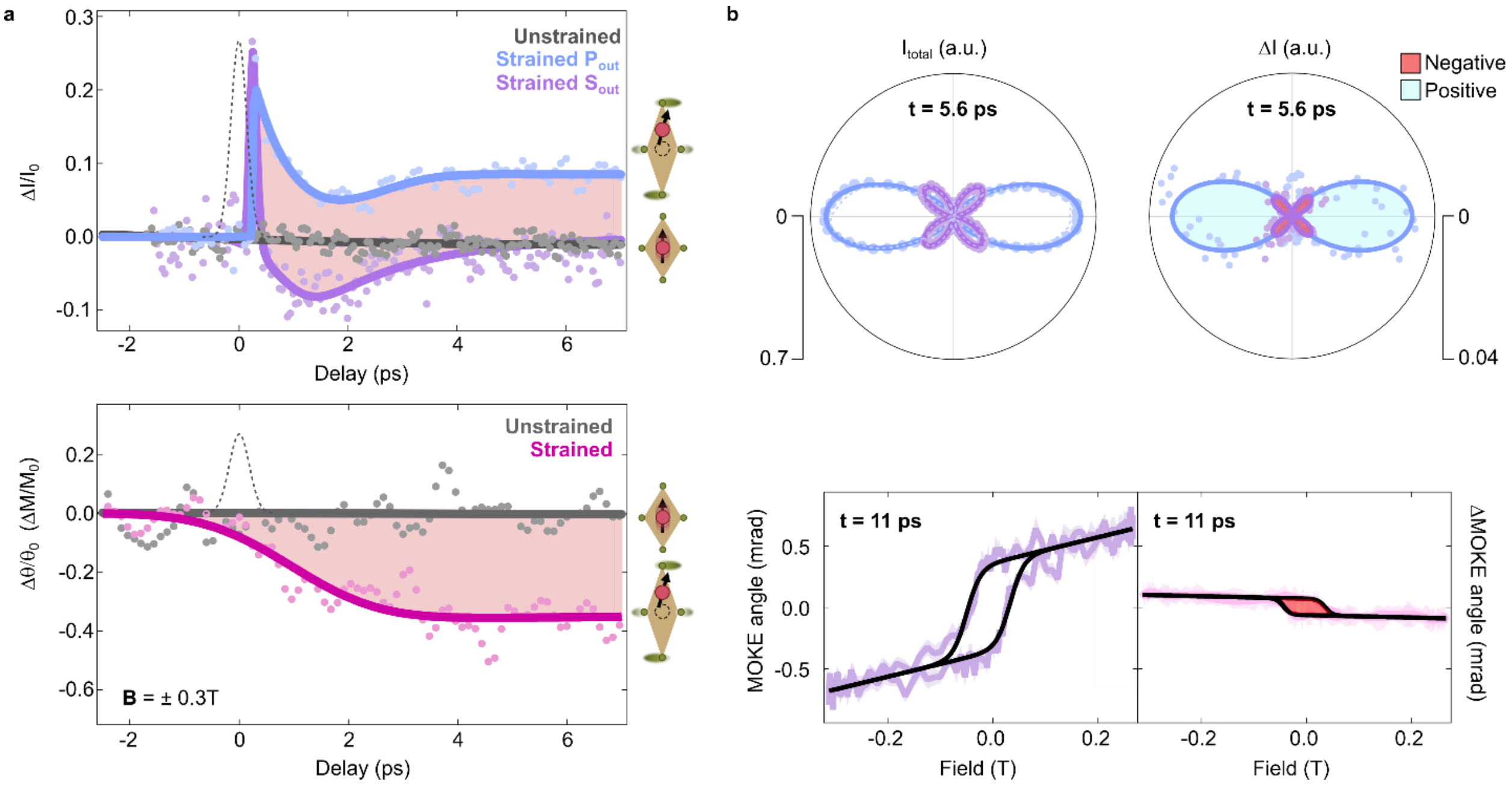

***Figure 3. Pump-induced SHG and MOKE dynamics. a)*** *Pump-induced dynamics of SHG (top) and MOKE signal (bottom), where the dashed lines in each plot indicate the pump pulse.* ***b)*** *Input polarization dependence of the SHG signal at 5.6 ps time delay after pump pulse (top), and field dependence of the MOKE signal at 11 ps time delay (bottom). Light, dashed lines in the pump-induced total SHG signal indicate the static SHG. For both SHG and MOKE data, the plots on the left show the total signal and plots on the right show only the change.*

Characterization of the symmetry and magnetism of the pump-induced state of the strained film is shown in Fig. 3b. In the long-lived state, the pump-induced SHG signal (upper panel) maintains the $C_{6v}$ symmetry of the equilibrium state, but with different dynamics for $P_{out}$ and $S_{out}$ channels—while at long times $P_{out}$ signal is enhanced, the $S_{out}$ signal is reduced. Corresponding pump-induced MOKE signals in the long-lived state show a clear magnetic-field hysteresis, which mirrors that found in the equilibrium film. Surprisingly, the pump-induced signal is found to have a lifetime >1 ns (see Supplementary information).

The distinct time dependences of the $P_{out}$ and $S_{out}$ channels of the SHG signal (Fig. 3a, blue and purple) suggest that the three independent $\chi^{(2)}$ tensor elements associated with the $C_{6v}$ symmetry ($d_{15}$, $d_{31}$, $d_{33}$) have different dynamics under resonant excitation. Changes in the relative ratio of the tensor elements indicate distortions of the unit cell. To gain further insight, we analyze the SHG response within a bond hyperpolarizability framework[42,43] (see Supplementary Information) and find that the experimentally measured $d_{33}/d_{31}$ ratios are consistent with $A_{2u}$ soft-mode distortions associated with the TBP $Fe^{3+}$ off centering. While quantitative extraction of the displacements requires further time-resolved x-ray measurements or theoretical modeling, the results establish that the pump excitation couples to the ferroelectric soft-mode, pushing the system towards a long-lived multiferroic state.

To verify the phononic origin of these effects, we examine the pump frequency dependence of the SHG signal, shown in Fig. 4a. As discussed above, pumping at 17.4 THz produces large relative changes in the SHG intensity, up to 20% (Fig. 4a, top). On the other hand, off-resonant excitation at 21 THz yields no detectable response (Fig. 4a, bottom). Furthermore, the frequency dependence of the pump-induced second harmonic signal closely follows the imaginary part of the dielectric function associated with the $E_{1u}$ phonon response (Fig. 4a, center inset)[22].

From these results, we can understand the dynamics of the polarization by considering the nonlinear phonon couplings in the strained film. Fitting the SHG response reveals two components, as noted above: a fast increase and decay within 2 ps after the pump, followed by a slow build up and saturation (Fig. 4a, top). The fast response follows the lifetime of the pumped IR mode, corresponding to a coupling between $Q_{IR}$ and the polarization ($P$) of the form $Q_{IR}^2 P$. The slow response, on the other hand, shows a rise time consistent with acoustic propagation through the 25 nm film, indicative of strain generation through the $Q_{IR}^2 \varepsilon$ coupling term discussed above. As no SHG response is observed in the unstrained sample, the electrostrictive-type coupling $P^2 \varepsilon$ cannot be responsible for the long-timescale polarization. Instead, a piezoelectric-like coupling of the form $P \cdot \varepsilon$ is necessary, enabled by the simultaneous coupling of the IR mode to the epitaxially induced strain and polarization (Fig. 4b). Consistent with this picture, oscillations centered at 1.5 THz corresponding to the polar soft mode[22,29] are observed in the SHG response on top of the strain signal (Fig. 4a, top inset). Thus, the $Q_{IR}^2 P$ and $P \cdot \varepsilon$ couplings that underlie the pump-enhanced ferroelectric polarization become allowed only in the presence of the *a priori* local symmetry breaking induced by epitaxial strain, consistent with recent theoretical work explaining the origin of long-lived pump-induced states[44].

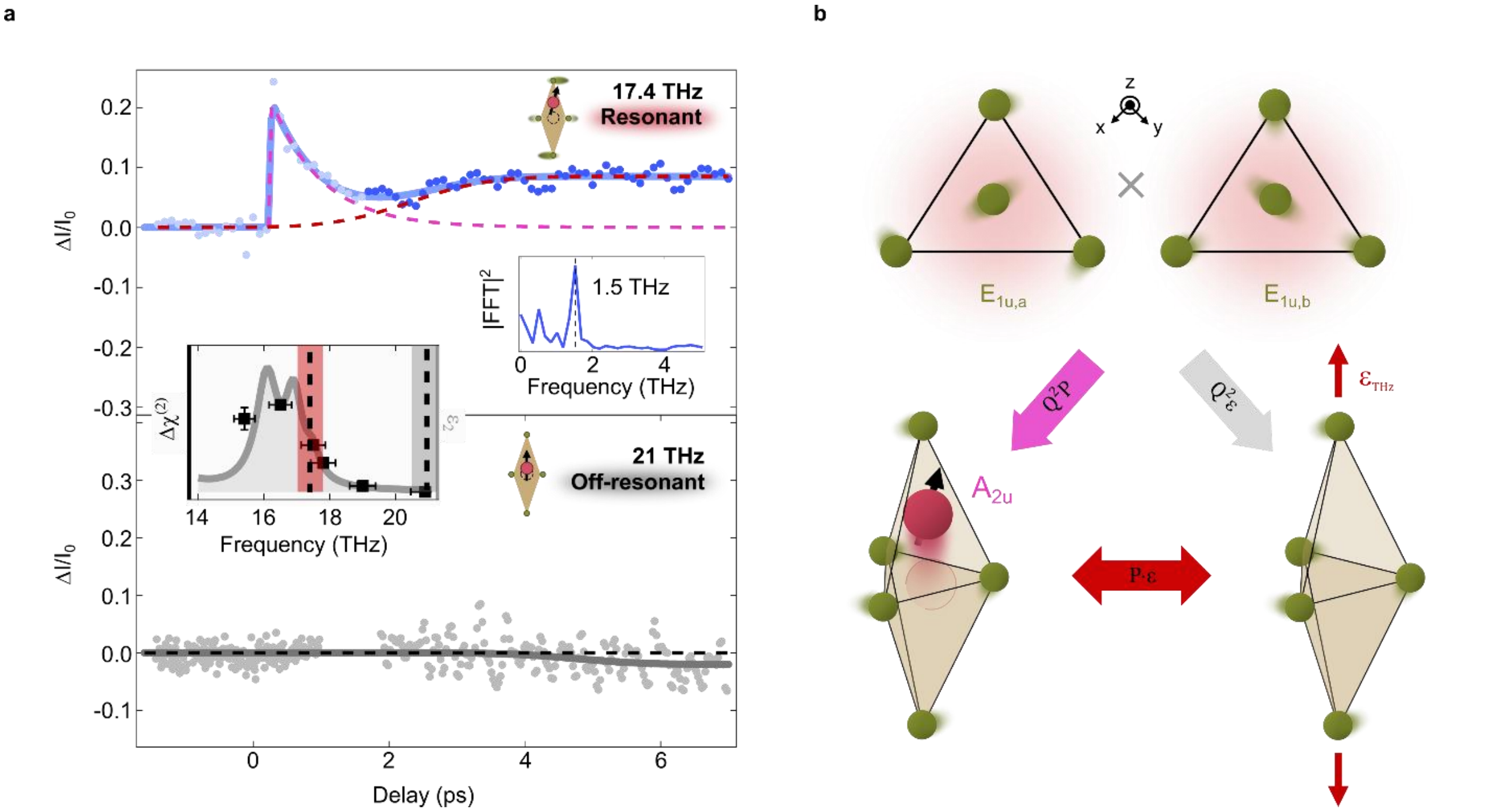

***Figure 4. Phononic origin of the pump-induced SHG signal. a)*** *top panel shows the change in the SHG signal when pumped resonantly at 17.4 THz, while the bottom panel shows the change when pumped off-resonantly at 21 THz. The inset shows the pump frequency dependence of the SHG signal and imaginary part of the dielectric function.* ***b)*** *Eigenmodes of the 17.4 THz $E_{1u}$ phonon mode (top) and the coupled strain along the c-axis (bottom, left) and increase in polarization corresponding to the $A_{2u}$ soft mode (bottom, right).* ***c)*** *Simulated signals of the strain and polarization due to resonant excitation at 17.4 THz.*

An additional mechanism is required to understand the magnetization dynamics upon pumping, as the remarkably large changes (up to ~40%) cannot be explained by thermal or electronic effects (see Supplementary Information). In particular, due to the multiferroic nature of the pump-induced state, we consider the coupling pathways between the driven IR mode, polarization, and magnetization. We measure the fluence dependences of the pump-induced SHG ($\Delta I/I_0$) and MOKE ($\Delta\theta/\theta_0$) responses at different times (Fig. 5a). Both the SHG and MOKE responses exhibit a combination of linear and quadratic fluence dependence, suggesting a possible coupling between the pump-induced structural distortion and the magnetic degrees of freedom. If no such coupling is present, we would expect the MOKE response to have a linear or sub-linear fluence dependence for a typical phonon-induced modification in magnetization, exchange interaction, or magnetic anisotropy[45-48]. Similarly, we would expect the pump-induced SHG to have linear and quadratic components (due to homodyning with the static SHG signal), the different dependences of $\Delta I/I_0$ for short and late time delays imply a different mechanism at play. Thus, the unusual fluence dependence seen in both the SHG and MOKE response motivates the development of a new theoretical framework that can account for these non-equilibrium interactions.

Strain-enabled bilinear coupling in modified ground state

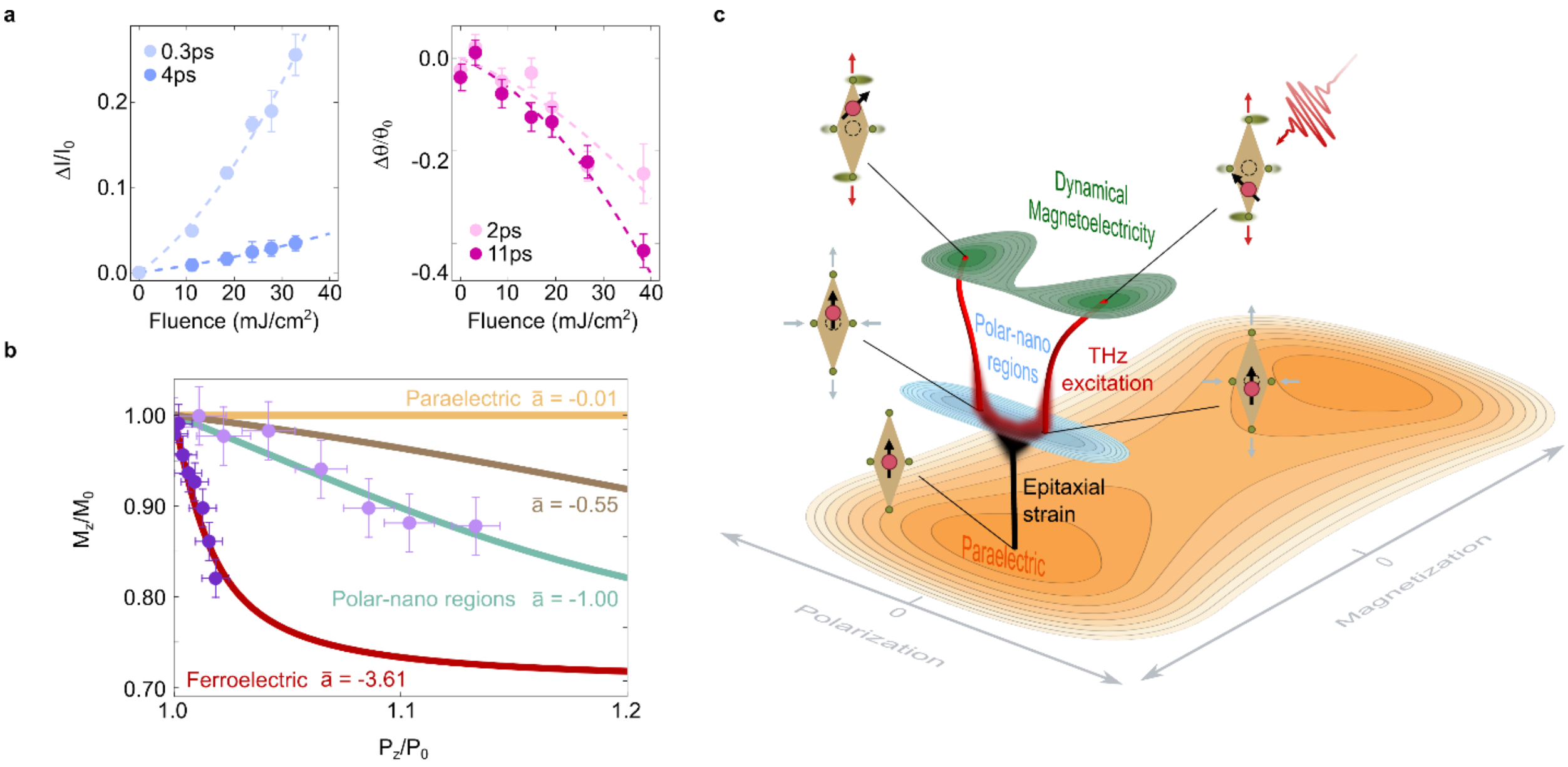

***Figure 5. Light-induced dynamical magnetoelectric coupling enabled via strain. a)*** *Pump fluence dependence of the SHG signal (left) and the MOKE signal (right). Both show a mixed linear-quadratic dependence, corroborating the bilinear coupling between polarization and magnetization.* ***b)*** *Parametric plot of SHG intensity data and MOKE rotation data (Fig. 5a) for early (dark purple) and late (light purple) time delays.* ***c)*** *Schematic of the free-energy landscape of the paraelectric, epitaxially strained, and resonantly excited states. The bilinear magnetoelectric coupling becomes accessible only in the strained state under resonant MIR excitation.*

The broken symmetry introduced by epitaxial strain enables coupling channels that are forbidden in the unstrained crystal, providing a natural pathway for the dynamical magnetoelectric interaction observed here. We start by expanding the free energy of the parent paramagnetic and paraelectric phase of BaM ($P6_3mc$ space group) in terms of the $c$-axis-aligned polarization $P_z$, easy-axis magnetization $M_z$, and the in-plane polarized two-dimensional IR-active phonon $\boldsymbol{Q} = (Q_x, Q_y)$ The lowest order term coupling these components, which satisfies the constraints within the allowed symmetries of the crystal has the form, $\Delta\Phi = \eta M_z P_z (\boldsymbol{M}_\perp \cdot \boldsymbol{Q})$ where $\boldsymbol{M}_\perp = (M_x, M_y)$ characterizes the in-plane components of the magnetization. In the bulk, below 720 K, the magnetic order parameter $M_z$ condenses while the system remains paraelectric ($P_z = 0$). In this case, $\Delta\Phi = 0$, and no phonon-driven effect should be present, as we observe in our experimental measurements. When compressive strain is applied in our epitaxial film, a weak local $P_z$ is established, allowing $\Delta\Phi$ to be non-zero. In this case, the quadrilinear free energy term collapses to a bilinear coupling, $\Delta\Phi = \eta' M_z P_z$, where the coupling coefficient is $\eta' = \eta\,(\boldsymbol{M}_\perp \cdot \boldsymbol{Q})$. This form suggests an effective coupling between the $z$-axis magnetization and polarization which can be driven by activating the IR phonons: a dynamical magnetoelectric effect.

In $BiFeO_3$, one of a few known room-temperature multiferroics, a static magnetoelectric coupling originates from the Dzyaloshinskii-Moriya interaction, which induces a canting whose strength is fixed by spin-orbit coupling. Here, the new dynamical coupling channel plays the same role, where the canting arises through a coupling between the in-plane components of $\boldsymbol{M}$ and $\boldsymbol{Q}$ and is therefore tunable with the phonon amplitude. A full analysis of the free energy expansion can be found in the Supplementary Information.

To verify this model, we plot the experimentally measured total magnetization ($M_z/M_0$) against the total polarization ($P_z/P_0$) as a function of fluence, up to 35 mJ/cm$^2$ (Fig. 5b, purple markers). We see a linear correlation, where the enhancement of $P_z$ corresponds to a linearly decreasing $M_z$, consistent for both short- and long-time delays (light and dark purple, respectively). We compare this correlation to the expected dynamical magnetoelectric coupling from the model described above for different paraelectric and ferroelectric initial states (Fig. 5b, solid lines). We parameterize the initial polarization with a dimensionless mode stiffness, $|\bar{\alpha}| = \sqrt{2\Delta E_P/\Delta E_K}$, which characterizes the ratio of the ferroelectric polarization barrier height $\Delta E_p$ to the magnetic anisotropy energy $\Delta E_K$. Using known values for the magnetic anisotropy energy and the best fit value of $\bar{\alpha} \approx -1$, we find that $\Delta E_p \approx 0.72$ meV at early time-delays (Supplementary Information). Comparing $\Delta E_p$ to the estimated inter-site dipole interaction energy (~ 1 meV)[26] verifies that the system lies in a weak ferroelectric regime, consistent with the picture of polar nanoregions in epitaxially strained BaM.

At long time-delays, we see that the value of $|\bar{\alpha}|$ increases to ~3.6, corresponding to a polarization barrier height of $\Delta E_p \approx 9.3$ meV, nearly an order of magnitude larger than the inter-site dipole energy. A simple interpretation of this behavior is that, at early time delays following phonon excitation, the local polarization is enhanced, but only short-range correlations are present, and the system remains in the disordered polar state formed by epitaxial strain. At long times, the build-up of light-induced strain introduces the dynamical magnetoelectric coupling that modifies the free energy landscape (as in Fig. 4b), driving the system into a long-range ordered ferroelectric state with strongly coupled ferroelectric and magnetic degrees of freedom.

Fig. 5c illustrates the evolution of the free energy landscape in the polarization-magnetization parameter space. The system initially resides in the fully relaxed paraelectric and ferrimagnetic equilibrium bulk state (orange layer). As epitaxial strain is introduced, it enters a weakly polarized regime characterized by shallow polarization wells leading to the formation of polar nanoregions (light blue layer). Resonant excitation with mid-infrared light further modifies the landscape, causing the polarization wells to deepen and move farther apart in polarization space. Through the bilinear coupling between polarization and magnetization, this enhanced polarization is accompanied by a gradual bending of the trajectory toward lower magnetization (red lines). This dynamically driven state maintains a strong magnetization and polarization, along with phonon-mediated magnetoelectric coupling, suggesting the stabilization of a light-induced multiferroic phase at room temperature (green layer).

We note the model presented here does not accurately describe the individual trajectories of the SHG and MOKE response or the behavior at high fluences, where biquadratic couplings of the IR mode to the polarization and magnetization, as well as higher-order anharmonic effects of the magnetoelectric coupling become important. In the Supplementary Information, we present an extended model with additional nonlinear effects that quantitatively fits the full range of experimental data. The results of the extended model agree with the conclusions determined from the simplified model above.

**Conclusions**

We have demonstrated that epitaxial strain enables a light-induced dynamical magnetoelectric coupling in BaM by generating polar nanoregions that break the local symmetry. Resonant driving of the $E_{1u}$ phonon activates an anharmonic coupling pathway that leads to long-ranged ferroelectric order and substantial modification of magnetization—an effect inaccessible in unstrained films or under static conditions. This magnetoelectric effect is dynamically tunable and enables a long-lived, pump-induced multiferroic state at room temperature. These results establish strain as a powerful tool for enabling non-equilibrium functionalities in materials that lack such coupling in equilibrium.

Future experiments could explore uniaxial strain, alternative substrates, or strain-gradient engineering to further tune the symmetry and strength of the coupling. Time-resolved x-ray scattering and ultrafast electron diffraction would allow direct tracking of the structure during resonant driving, enabling quantitative extraction of mode amplitudes and verification of the dynamical multiferroic pathway.

**References**

1 Först, M. et al. Nonlinear phononics as an ultrafast route to lattice control. *Nature Physics* **7**, 854–856 (2011).

2 Subedi, A., Cavalleri, A. & Georges, A. Theory of nonlinear phononics for coherent light control of solids. *Physical Review B* **89**, 220301 (2014).

3 Mankowsky, R. et al. Nonlinear lattice dynamics as a basis for enhanced superconductivity in $YBa_2Cu_3O_{6.5}$. *Nature* **516**, 71–73 (2014).

4 Disa, A. S., Nova, T. F. & Cavalleri, A. Engineering crystal structures with light. *Nature Physics* **17**, 1087–1092 (2021).

5 Kampfrath, T., Tanaka, K. & Nelson, K. A. Resonant and nonresonant control over matter and light by intense terahertz transients. *Nature Photonics* **7**, 680–690 (2013).

6 Dhillon, S. S. et al. The 2017 terahertz science and technology roadmap. *Journal of Physics D: Applied Physics* **50**, 043001 (2017).

7 Song, L. et al. Review of intense terahertz generation and its emerging applications. *Applied Physics Reviews* **13**, 021302 (2026).

8 Juraschek, D., Fechner, M. & Spaldin, N. Ultrafast Structure Switching through Nonlinear Phononics. *Physical Review Letters* **118**, 054101 (2017).

9 Hortensius, J. R., Afanasiev, D., Sasani, A., Bousquet, E. & Caviglia, A. D. Ultrafast strain engineering and coherent structural dynamics from resonantly driven optical phonons in $LaAlO_3$. *npj Quantum Material*s **5**, 95 (2020).

10 Liu, B. et al. Generation of narrowband, high-intensity, carrier-envelope phase-stable pulses tunable between 4 and 18 THz. *Optics Letters* **42**, 129 (2017).

11 Nova, T. F., Disa, A. S., Fechner, M. & Cavalleri, A. Metastable ferroelectricity in optically strained $SrTiO_3$. *Science* **364**, 1075–1079 (2019).

12 Li, X. et al. Terahertz field–induced ferroelectricity in quantum paraelectric $SrTiO_3$. *Science* **364**, 1079–1082 (2019).

13 Rini, M. et al. Control of the electronic phase of a manganite by mode-selective vibrational excitation. *Nature* **449**, 72–74 (2007).

14 Disa, A. S. et al. Polarizing an antiferromagnet by optical engineering of the crystal field. *Nature Physics* **16**, 937–941 (2020).

15 Disa, A. S. et al. Photo-induced high-temperature ferromagnetism in $YTiO_3$. *Nature* **617**, 73–78 (2023).

16 Maehrlein, S. F. et al. Dissecting spin-phonon equilibration in ferrimagnetic insulators by ultrafast lattice excitation. *Science Advances* **4**, eaar5164 (2018).

17 Stupakiewicz, A. et al. Ultrafast phononic switching of magnetization. *Nature Physics* **17**, 489–492 (2021).

18 Basini, M. et al. Terahertz electric-field-driven dynamical multiferroicity in $SrTiO_3$. *Nature* **628**, 534–539 (2024).

19 Kumar, S., Prakash, J., Verma, A. & Jasrotia, R. A comprehensive review of synthesis, properties, and applications of $BaFe_{12}O_{19}$ hexaferrites. *Materials Today: Proceedings* S2214785323009100 (2023).

20 Shen, S.-P. et al. Magnetic-ion-induced displacive electric polarization in FeO5 bipyramidal units of $(Ba,Sr)Fe_{12}O_{19}$ hexaferrites. *Physical Review B* **90**, 180404 (2014).

21 Rowley, S. E. et al. Uniaxial ferroelectric quantum criticality in multiferroic hexaferrites $BaFe_{12}O_{19}$ and $SrFe_{12}O_{19}$. *Scientific Reports* **6**, 25724 (2016).

22 Mikheykin, A. S. et al. Lattice anharmonicity and polar soft mode in ferrimagnetic M-type hexaferrite $BaFe_{12}O_{19}$ single crystal. *The European Physical Journal* B **87**, 232 (2014).

23 Pullar, R. C. Hexagonal ferrites: A review of the synthesis, properties and applications of hexaferrite ceramics. *Progress in Materials Science* **57**, 1191–1334 (2012).

24 Manglam, M. K., Kumari, S., Mallick, J. & Kar, M. Crystal structure and magnetic properties study on barium hexaferrite of different average crystallite size. *Applied Physics A* **127**, 138 (2021).

25 Marshall, S. P. & Sokoloff, J. B. Phonon spectrum for barium ferrite. *Physical Review B* **44**, 619–627 (1991).

26 Zhang, X., Ye, Q.-J., Xiang, H. & Li, X.-Z. Quantum paraelectricity of $BaFe_{12}O_{19}$. *Physical Review B* **101**, 104102 (2020).

27 Shen, S.-P. et al. Quantum electric-dipole liquid on a triangular lattice. *Nature Communications* **7**, 10569 (2016).

28 Cao, H. B. et al. High pressure floating zone growth and structural properties of ferrimagnetic quantum paraelectric BaFe12 O19. *APL Materials* **3**, 062512 (2015).

29 Wang, P. & Xiang, H. Room-Temperature Ferrimagnet with Frustrated Antiferroelectricity: Promising Candidate Toward Multiple-State Memory. *Physical Review X* **4**, 011035 (2014).

30 Waqar, M. et al. Synthesis and properties of nickel-doped nanocrystalline barium hexaferrite ceramic materials. *Applied Physics A* **124**, 286 (2018).

31 Shao, Y. et al. Multi-susceptible single-phase $BaAl_xFe_{12-x}O_{19}$ ceramics with both improved magnetic and ferroelectric properties. *Applied Physics Letters* **114**, 242902 (2019).

32 Apostolova, I. N., Apostolov, A. T., Trimper, S. & Wesselinowa, J. M. Multiferroic Properties of Pure, Transition-Metal and Rare-Earth–Doped $BaFe_{12}O_{19}$ Nanoparticles. *Physica status solidi (b)* **258**, 2100069 (2021).

33 Rafiq, M. A., Waqar, M., Mirza, T. A., Farooq, A. & Zulfiqar, A. Effect of $Ni^{2+}$ Substitution on the Structural, Magnetic, and Dielectric Properties of Barium Hexagonal Ferrites ($BaFe_{12}O_{19}$). *Journal of Electronic Materials* **46**, 241–246 (2017).

34 Guguschev, C. et al. Revisiting the Growth of Large (Mg,Zr):$SrGa_{12}O_{19}$ Single Crystals: Core Formation and Its Impact on Structural Homogeneity Revealed by Correlative X-ray Imaging. *Crystal Growth & Design* **22**, 2557–2568 (2022).

35 K. P., H. et al. Polar Nano-Domains in Barium Hexaferrite Revealed with Multislice Electron Ptychography. *Microscopy and Microanalysis* **28**, 476–478 (2022).

36 Harikrishnan, K. P. et al. Visualizing Polar Distortions and Interface Effects with Multislice Ptychography. *Microscopy and Microanalysis* **29**, 1626–1627 (2023).

37 Y. E. Li et al., Strain-Induced Relaxor Multiferroicity at Room Temperature in Hexaferrite $BaFe_{12}O_{19}$ Thin Films. Preprint at arXiv:2609.21352.

38 Gerber, R., Atkinson, R. & ˇSimˇsa, Z. Magnetism and magneto-optics of hexaferrite layers. *Journal of Magnetism and Magnetic Materials* **175**, 79–89 (1997).

39 Atkinson, R., Papakonstantinou, P., Salter, I. & Gerber, R. Optical and magneto-optical properties of Co-Ti-substituted barium hexaferrite single crystals and thin films produced by laser ablation deposition. *Journal of Magnetism and Magnetic Materials* **138**, 222–231 (1994).

40 Jakubisová, E. et al. Magneto-optical studies of $BaFe_{12}O_{19}$ films grown by metallo-organic decomposition. Optical Materials Express **5**, 1323 (2015).

41 Fechner, M. *et al.* Quenched lattice fluctuations in optically driven $SrTiO_3$. *Nat. Mater.* **23**, 363–368 (2024).

42 Powell, G. D., Wang, J.-F. & Aspnes, D. E. Simplified bond-hyperpolarizability model of second harmonic generation. *Physical Review B* **65**, 205320 (2002).

43 Xue, D. & Zhang, S. Bond-charge calculation of nonlinear optical susceptibilities of $LiXO_3$ type complex crystals. *Chemical Physics* **226**, 307–318 (1998).

44 Li, S. J., Khalsa, G., Kaaret, J. Z. & Benedek, N. A. Understanding long-lived metastable phases in ultrafast optical experiments. *npj Quantum Mater.* **11**, 2 (2025).

45 Nova, T. F. et al. An effective magnetic field from optically driven phonons. *Nature Physics* **13**, 132–136 (2017).

46 Afanasiev, D. et al. Ultrafast control of magnetic interactions via light-driven phonons. *Nature Materials* **20**, 607–611 (2021).

47 Fechner, M. et al. Magnetophononics: Ultrafast spin control through the lattice. *Physical Review Materials* **2**, 064401 (2018).

48 Gu, M. & Rondinelli, J. M. Nonlinear phononic control and emergent magnetism in Mott insulating titanates. *Physical Review B* **98**, 024102 (2018).

## Methods

<u>Sample preparation</u>

Thin films of $BaFe_{12}O_{19}$ were grown by reactive-oxide molecular-beam epitaxy (MBE) in a Veeco GEN10 system. (0001) sapphire substrates were annealed in air at 1000 °C for 6 h in a tube furnace and subsequently used as calibration substrates to establish the optimal growth conditions. As-polished (0001) hexaferrite substrates $Sr_{1.03}Ga_{10.81}Mg_{0.58}Zr_{0.58}O_{19}$ (SGMZ), were subsequently used for the growth of fully strained $BaFe_{12}O_{19}$ films. To ensure iron is oxidized into the +3 valent state, distilled ozone (~ 80% $O_3$ + 20% $O_2$ reaching the substrate) was used as the oxidant and the background pressure of it was kept at $1.0 \times 10^{-6}$ torr during growth. $BaFe_{12}O_{19}$ was grown by co-deposition of barium and iron at a substrate temperature of 750 °C measured by an optical pyrometer operating at a wavelength of 1550 nm. Precise stoichiometry control is essential for achieving a fully strained film. A method utilizing in situ reflection high-energy electron diffraction (RHEED), X-ray diffraction (XRD), and atomic force microscopy (AFM) to determine the optimal iron-to-barium flux ratios is described in Ref. 37. RHEED patterns were recorded using the KSA-400 software and a Staib electron source operated at 13 kV and a filament current of 1.5 A. XRD scans were measured with a PANalytical Empyrean diffractometer with Cu Kα1 radiation. AFM surface topography images were taken with the Asylum Research Cypher ES Environmental AFM.

Experimental design for TR-MOKE and TR-SHG

In order to generate the required 17.4 THz pump pulse for resonant excitation of the $E_u$ lattice distortion modes, we used two signal outputs of the optical parametric amplifier, centered at 1.42 um and 1.53 um respectively, for a difference frequency generation (DFG) process in the inorganic crystal GaSe. Each OPA signal was pumped by the output of a Ti:Sapphire amplifier at 800 nm, 1 kHz repetition rate with 35 fs pulses. To produce narrow-band THz pulses, we used two pairs of gratings to introduce GDD to each OPA output and stretch the pulses to 400 fs before combining on a 1 mm GaSe crystal. The resulting THz frequency and electric field was measured by electro-optic sampling via another 30 um GaSe crystal at the sample position. For all TR-MOKE and TR-SHG experiments, the pump durations and bandwidths were around 340-505 fs and 1.5-1.8 THz. Typical pump spot sizes ranged between 38 um to 55 um for SHG and MOKE experiments.

For the Kerr rotation measurements, a polar MOKE configuration was used with the applied magnetic field perpendicular to the sample surface and parallel to the plane of incidence. Both the pump and probe pulses were at normal incidence, which was done by using an off-axis parabolic mirror with a center hole to allow collinear probe propagation. The magnetic field was applied by using opposite polar magnets placed directly behind the sample, allowing a continuous change in the magnetic field by moving the magnets longitudinally. The applied field at the sample position was measured by a Hall probe centered at the focus position of the pump/probe pulses. The probe pulse with a center wavelength of 400 nm was generated via SHG through a BBO crystal, with its polarization axis set by a polarizer before the sample. The incident probe polarization was s-polarized, perpendicular to the THz pump polarization to eliminate contributions from THz-driven birefringence. After reflecting from the sample surface, the probe beam was routed to two Si photodetectors, after passing through a half-wave plate (HWP) and a polarizing Wollaston prism. The two orthogonal components coming out of the Wollaston prism were balanced by the HWP such that the photodetector outputs were zero before each static and pump-probe experiment. Equilibrium measurements for both MOKE and SHG were performed in the same experimental setup as the pump-probe experiments, with the pump blocked.

For second harmonic generation (SHG) measurements, the same pump configuration was employed using an off-axis parabolic mirror. To probe the *c*-axis mirror-plane symmetry breaking expected to accompany the onset of ferroelectricity, the probe beam was incident at 45° relative to the a–b plane of the crystal. Rotational anisotropy measurements were performed on the reflected SHG signal by rotating the incident probe polarization with a half-wave plate, where P-polarization was defined to be parallel to the a or b crystallographic axis. The horizontally and vertically polarized SHG components were detected independently using a photomultiplier tube for both static and pump–probe measurements. In pump-probe experiments, narrow bandwidth band-pass filters centered at 808 nm and 405 nm were used to suppress the $\chi^{(3)}$ contribution from the sample.

<u>SHG signal analysis</u>

The unstrained centrosymmetric sample belongs to the point group $D_{6h}$. This structural distortion associated with the biaxial compression in the films is reflected in the emergence of a finite second harmonic generation (SHG) signal, with rotational anisotropy measurements consistent with $C_{6v}$ symmetry. The second-order nonlinear polarization induced by the probe electric field is given by

$$P_i(2\omega) = \epsilon_0 \left( \sum_{jk} \chi_{ijk}^{(2)} E_j(\omega) E_k(\omega) \right)$$

Where $\chi_{ijk}^{(2)}$ is the second-order nonlinear susceptibility tensor and $E_j(\omega)$ are the components of the incident probe electric field. For an incident angle $\theta$ and probe polarization angle $\phi$, the electric field is written as

$$E(\omega) = E_0 \begin{pmatrix} \sin\phi\cos\theta \\ \cos\phi \\ \sin\phi\sin\theta \end{pmatrix}$$

For a crystal with $C_{6v}$ symmetry, the nonlinear susceptibility tensor may be written in contracted notation as

$$d = \begin{pmatrix} 0 & 0 & 0 & 0 & d_{15} & 0 \\ 0 & 0 & 0 & d_{15} & 0 & 0 \\ d_{31} & d_{31} & d_{33} & 0 & 0 & 0 \end{pmatrix}$$

where the independent tensor elements are $d_{15}$, $d_{31}$, $d_{33}$.

The horizontal and vertical SHG components were fit independently using the expressions:

$$I(\phi)_H = \left[\frac{1}{2}(d_{15} + d_{31} + d_{33})\sin^2\phi + d_{33}\cos^2\phi\right]^2$$

$$I(\phi)_V = 2d_{15}^2 \sin^2\phi \cos^2\phi$$

These measurements allow the independent extraction of the nonlinear susceptibility tensor elements. In the Kleinman symmetry limit, one expects $d_{15} = d_{31}$; however, the strained films exhibit a clear deviation from this condition, with $d_{15} \neq d_{31}$.

For the time-resolved measurements, polarization-dependent data were collected at three representative delay times: 90 fs, 0.3 ps, and 5.6 ps. Our bond hyperpolarizability analysis, described in the Supplementary Material, focuses on the long-timescale response (5.6 ps delay time). At this timescale, the strain predominantly modifies the $d_{33}$ and $d_{31}$ tensor elements, both of which increase, as evidenced by the enhanced horizontally polarized SHG signal. In contrast, the $d_{15}$ element, governing the vertically polarized response, exhibits a comparatively small decrease.

MOKE signal analysis

For static MOKE measurements, the polarization rotation angle was calculated by measuring the difference in the signal intensity of the two orthogonal components, $I_s$ and $I_p$ for S and P polarizations:

$$\theta = \frac{\Delta I}{2I_{\text{sum}}} \tag{1}$$

where $\Delta I = I_s - I_p$ and $I_{sum} = I_s + I_p$. The rotation angle can be expressed in terms of the magneto-optical coefficient and the magnetization of the sample by using the Jones matrix formalism, starting with the sample matrix

$$S = \begin{pmatrix} \tilde{r}_{pp} & \tilde{r}_{ps} \\ \tilde{r}_{sp} & \tilde{r}_{ss} \end{pmatrix}. \tag{2}$$

$\tilde{r}_{ij}$ are the complex Fresnel coefficients representing the ratio of the $i$-polarized component of the reflected light and the $j$-polarized component of the incident light. The sample matrix can be written in another form that has explicit magnetization dependence,

$$S = \begin{pmatrix} r & r(\beta' + i\beta'')M \\ -r(\beta' + i\beta'')M & r \end{pmatrix}. \tag{3}$$

Here $\beta'$ and $\beta''$ are the real and imaginary parts of the magneto-optical coefficient. Using this form of the sample matrix along with the Jones matrices for the incident light and the half-wave plate defined as

$$E = \begin{pmatrix} E_p \\ E_s \end{pmatrix} \tag{4}$$

and

$$H = \begin{pmatrix} \cos 2\alpha & \sin 2\alpha \\ \sin 2\alpha & -\cos 2\alpha \end{pmatrix} \tag{5}$$

where $\alpha$ is the angle of the half-wave plate with respect to the horizontal. The final output intensities of the sum and difference signals at $\alpha = \frac{\pi}{8}$ are

$$I_{\text{sum}} = r^2\left(1 + M^2(\beta'^2 + \beta''^2)\right) \tag{6}$$

$$\Delta I = 2Mr^2\beta'. \tag{7}$$

Then, by evaluating $\frac{\Delta I}{2I_{\text{sum}}}$, we can see that the rotation angle is a linear function of M

$$\theta = \frac{\Delta I}{2I_{sum}} = \frac{\beta' M}{1 + (\beta'^2 + \beta''^2)M^2} \approx \beta' M. \tag{8}$$

Here we have assumed that the complex Kerr rotation angle $\Theta = (\beta' + i\beta'')M = \theta + i\varepsilon$ is small, where $\theta$ is the polarization rotation angle and $\varepsilon$ is the ellipticity. The pump-induced change in the polarization rotation is defined as

$$\Delta\theta(t) = \frac{\Delta\theta(+B,t) - \Delta\theta(-B,t)}{2} \tag{9}$$

which eliminates possible nonmagnetic effects, where $B$ is the applied magnetic field and $t$ is time. In order to determine the pump-induced magnetization of the sample, we can expand the static MOKE angle analysis where the magnetization in the static case is

$$M_{\text{static}} = \frac{\theta}{\beta'} \tag{10}$$

and thus the pump-induced magnetization becomes

$$M_{\text{pumped}}(t) = \frac{\theta + \Delta\theta(t)}{\beta'} \tag{11}$$

where $\theta$ and $\Delta\theta(t)$ are the static and pump-induced rotation angles as defined previously.

Distinguishing magnetic and optical effects in MOKE signals

To determine how much of the pump-induced change in the rotation angle is purely due to change in the internal magnetization of the sample, we expand the pump-induced change in the complex Kerr angle in terms of the differential changes of the magneto-optical coefficient and the magnetization

$$\Delta\Theta = M\Delta\beta + \beta\Delta M. \tag{12}$$

The complex magneto-optical coefficient can be expressed as $\beta = A\exp(i\phi)$ where $\phi$ is the phase. The complex Kerr angle then can be expressed in terms of relative variations $\delta\beta$, $\delta M$, and $\delta A\left(\delta\beta = \frac{\Delta\beta}{\beta}, \delta M = \frac{\Delta M}{M}, \delta A = \frac{\Delta A}{A}\right)$[49]:

$$\delta\Theta = \delta A + \delta M + i\Delta\phi. \tag{13}$$

The complex Kerr angle can also be expressed only in terms of the rotation and ellipticity:

$$\delta\Theta = \frac{\theta\delta\theta + i\epsilon\delta\epsilon}{\theta + i\epsilon}. \tag{14}$$

Equating the two expressions for the complex Kerr angle, we can solve for the relative changes in rotation and ellipticity

$$\delta\theta = \delta A + \delta M - \frac{\Delta\phi}{k}$$
$$\delta\epsilon = \delta A + \delta M + k\Delta\phi$$

where $k = \frac{\theta}{\epsilon}$. Then, we can define the relative change in magnetization as

$$\delta M = \frac{k^2\delta\theta + \delta\epsilon}{k^2 + 1} - \delta A. \tag{15}$$

Assuming that the dynamics of the real and imaginary parts of the change in magneto-optical coefficients are similar

$$\delta A = \gamma\Delta\phi \tag{16}$$

then the relative change in magnetization simplifies to

$$\delta M = \frac{\delta\theta + \alpha\delta\epsilon}{1 + \alpha} \tag{17}$$

where $\alpha = \frac{1-\gamma k}{k^2+\gamma k}$. Here we can see that if $\delta\theta = \delta\epsilon$, the relative change in magnetization is identical to the relative change in rotation $\delta M = \delta\theta$. For cases where the relative changes in the rotation and ellipticity are not equal, the simplification cannot be made and a more careful analysis must be done, which we explain further in the Supplementary Information.

<u>SHG vs. MOKE parametric curve</u>

Pump-probe SHG and MOKE data were interpolated at fixed fluence values (0 – 35 mJ/cm$^2$) in order to generate the parametric curve presented in Fig. 5b, as the two data sets were taken at different fluence values due to experimental constraints.

## References

49 Razdolski, I. et al. Analysis of the time-resolved magneto-optical Kerr effect for ultrafast magnetization dynamics in ferromagnetic thin films. *Journal of Physics: Condensed Matter* **29**, 174002 (2017).

## Data availability

The data supporting the findings of this study are available from the corresponding author upon reasonable request.

## Acknowledgements

We thank Professor Martin Dressel for providing room temperature BaM phonon parameters that were used to calculate the dielectric permittivity in this work. This work was supported by funding from the U.S. Department of Energy, Office of Science, Office of Basic Energy Sciences, under Contract No. DE-SC0019414. Y.E.L., H.K., D.A.M., and D.G.S. acknowledge support from the Army Research Office under the ETHOS MURI via cooperative agreement W911NF-21-2-0162. This research was funded in part by the Gordon and Betty Moore Foundation's EPiQS Initiative through Grants GBMF3850 and GBMF9073 to support the work of DGS. The authors acknowledge the Texas Advanced Computing Center (TACC) at The University of Texas at Austin for providing computational resources that have contributed to the research results reported within this paper.

## Author contributions

## Competing interests

The authors declare no competing interests.

## Additional information

Supplementary information is available for this paper at [link to online content].